\documentclass{appolb}
\usepackage{epsfig}
\usepackage{amsmath}
\usepackage{enumitem}
\usepackage{graphicx}
\usepackage{cite}
\usepackage{sidecap}
\usepackage[colorlinks=true,linkcolor=black,citecolor=black,urlcolor=black]{hyperref}

\begin{document}
\title{Overview of results from NA61/SHINE\thanks{Presented at the 
    XXXII Cracow Epiphany Conference on the recent results from Heavy Ion Physics,
    January 12-16, 2026, Krak\'ow, Poland.}
}
\author{Andrzej Rybicki$^1$
\vspace*{0.2cm}
for the NA61/SHINE collaboration
\address{$^1$Institute of Nuclear Physics, Polish 
 Academy of Sciences, \\
Radzikowskiego 152, 31-342~Krak\'ow, Poland}
}
\maketitle
\begin{abstract}
NA61/SHINE is a multipurpose, fixed-target spectrometer operating at the CERN SPS. The studied regime of collision energies, 5.1$<$$\sqrt{s_\mathrm{NN}}$$<$16.8 /27.4 GeV, places the project in-between the two main European heavy ion activities of the coming decade, the continued LHC (0.9$<$$\sqrt{s_\mathrm{NN}}$$<$14 TeV) and the announced FAIR SIS100 (2.7$<$$\sqrt{s_\mathrm{NN}}$$<$4.9 GeV) programs. Also, the project partially overlaps with RHIC BES and STAR-FXT (3$<$$\sqrt{s_\mathrm{NN}}$$<$ 62.4 GeV, with data taking completed). This contribution gives a subjective summary of the recent results from NA61/SHINE, with particular emphasis on these of greatest importance for the other research programs.
 \end{abstract}

DOI:


\section{Introduction}
\label{sec1}

The NA61/SHINE experiment started in 2007, as a direct inheritor of the NA49 experiment at the SPS     (the latter being inspired by the physical findings from NA35 at the same synchrotron). The experiment follows three complementary research programs: (a) studies of strong interactions (b)~reference measurements for JPARC and Fermilab neutrino programs, aimed at the reduction of systematic uncertainties of the neutrino flux, and (c)~measurements for cosmic-ray physics (Pierre-Auger, the Telescope Array, IceTop) aimed at refining air-shower simulations and a better understanding of
cosmic-ray propagation in the galaxy. With rare exceptions, only the first program~(a) will be addressed here; the interested reader is invited to consult Ref.~\cite{na61} for results from the other programs.

\begin{figure}[!h]
  \centering
  \hspace*{-3.8cm}\includegraphics[width=0.69\textwidth]{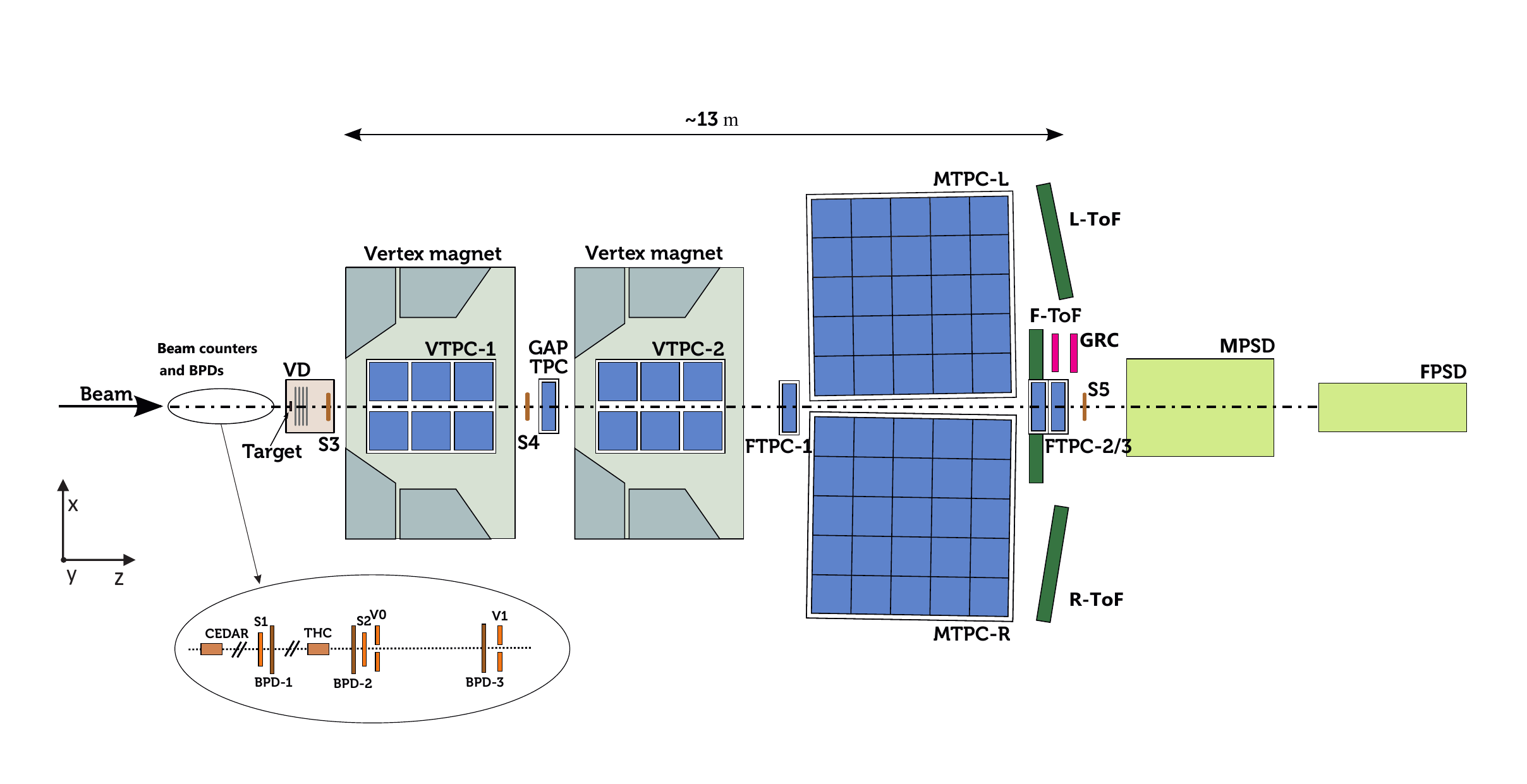}
  \begin{picture}(10,10)
\put(-246,0){\includegraphics[width=4.4cm,height=1.25cm]{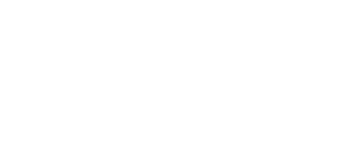}}
\put(-246,26){\includegraphics[width=0.7cm,height=1.16cm]{arblank.pdf}}
\put(-226,26){\includegraphics[width=0.48cm,height=1.1490cm]{arblank.pdf}}
\put(-215,26){\includegraphics[width=0.48cm,height=0.8990cm]{arblank.pdf}}
\hspace*{-0.0cm}
\put(6,18){\includegraphics[width=0.32\textwidth]{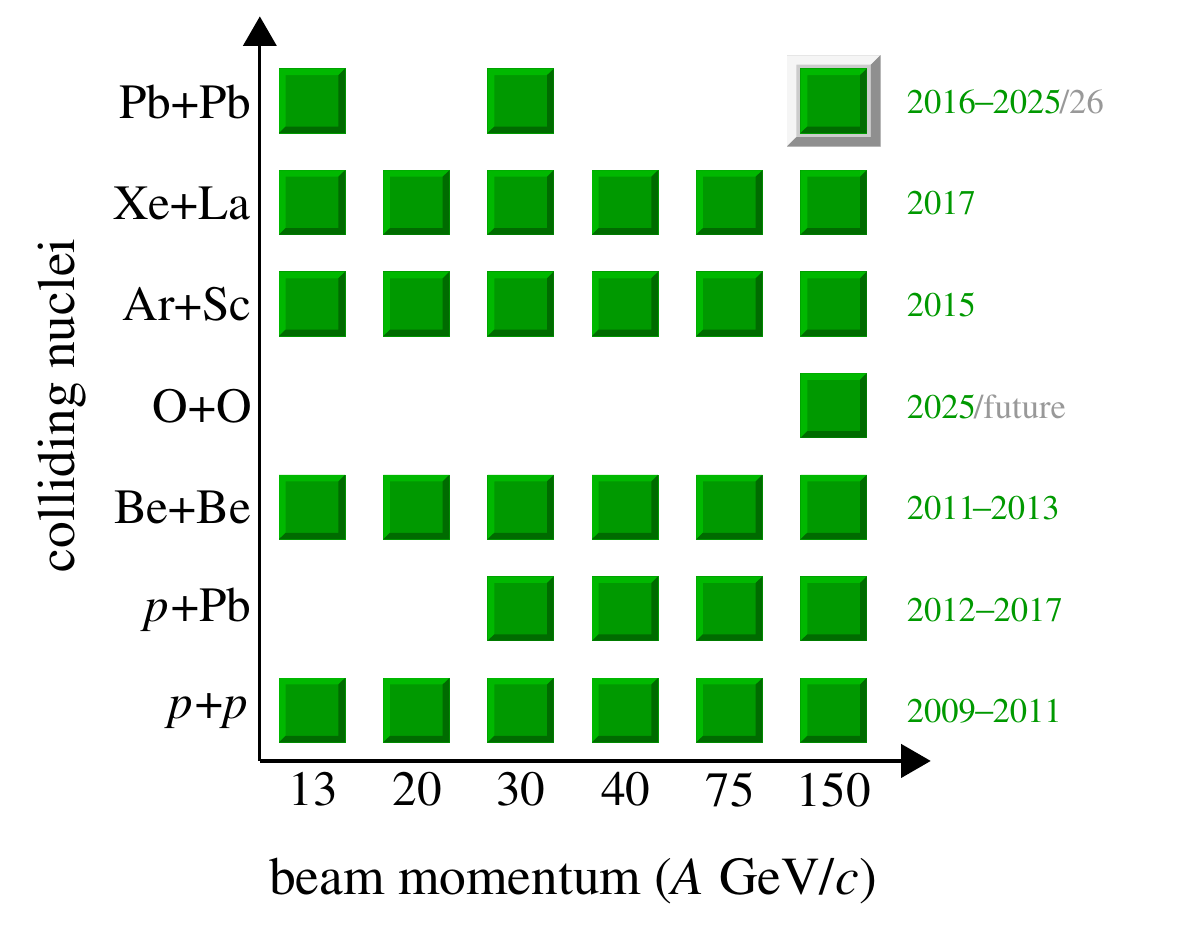}}
\end{picture}
\vspace*{-0.9cm}
\caption{Present layout of the NA61/SHINE detector (left) and dataset collected by the strong interaction program of the experiment (right).\vspace*{-1mm}}
   \label{f1}
\end{figure}

The fixed-target configuration of the experiment, very different from the collider-based configuration specific for most of LHC detectors, is evident in Fig.~\ref{f1} (left). The setup comprises 8 Time Projection Chambers (TPC) for charged particle tracking, momentum measurement and particle identification via ionization energy loss, supplemented with time-of-flight (ToF) detectors improving PID in limited areas of phase-space, and a forward calorimetry system for centrality triggering and definition through the measurement of the forward energy deposit (PSD). This layout ensures a set of advantages of NA61/SHINE as a hadron spectrometer, in comparison to collider experiments operating at the TeV scale:

\begin{itemize}
  \vspace*{-0.1mm}
  \item
 Thanks to its extended, asymmetric setup as well as the moderate collision energy (beam momenta up to 150$A$ and 400 GeV/\hspace*{-0.1mm}$c$ for nuclear and hadronic beams, respectively), the acceptance of the detector covers nearly the entire projectile c.m.s. hemisphere of the collisions for charged hadrons, plus a part of the target hemisphere for specific neutrals. Numerous NA61/SHINE measurements provide continuous coverage from mid- up to beam-rapidity.
  \vspace*{-0.1mm}
 \item
 The measurements do not suffer from the low-$p_T$ cutoff unavoidable for collider experiments. Typically, the obtained transverse momentum distributions start sharply at $p_T$=0.
\end{itemize}
  \vspace*{-0.0mm}
Thanks to the broad range of beam momenta extractable from the SPS and the practical accessibility of different types of fixed targets, the experiment accumulated a versatile, ``regular'' dataset of system-size versus collision energy configurations, allowing for a 2D scan of various characteristics of particle production (Fig.~\ref{f1}, right). Selected results will be presented below. 

  \vspace*{-0.1mm}
\section{Charged pion and kaon production per wounded nucleon}
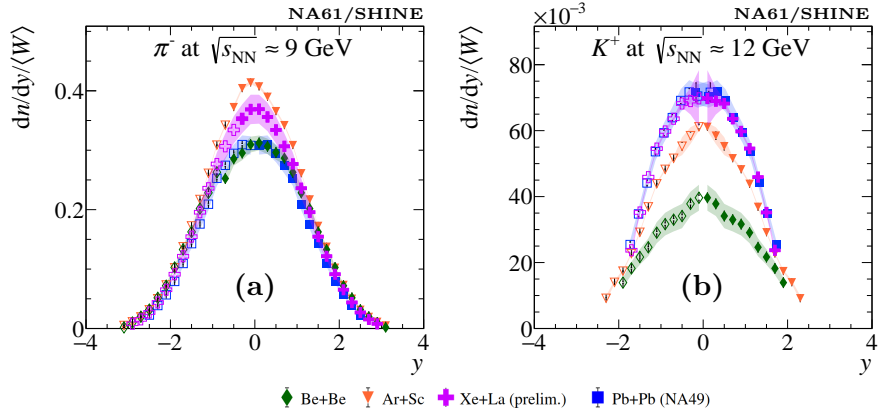
\begin{figure}[!h]
  \centering
  \hspace*{0cm}\includegraphics[width=0.458\textwidth]{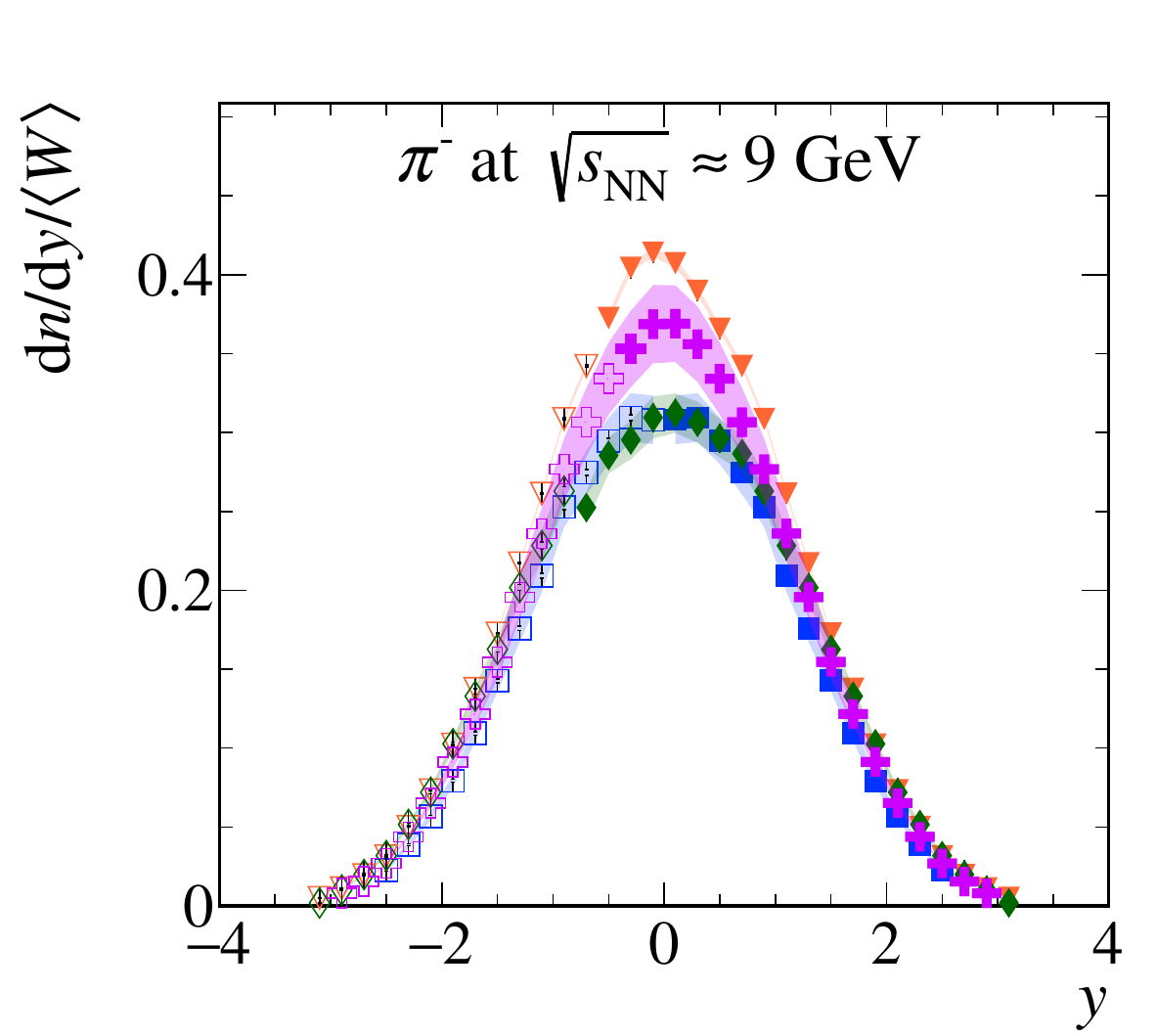}
  \hspace*{0cm}\includegraphics[width=0.458\textwidth]{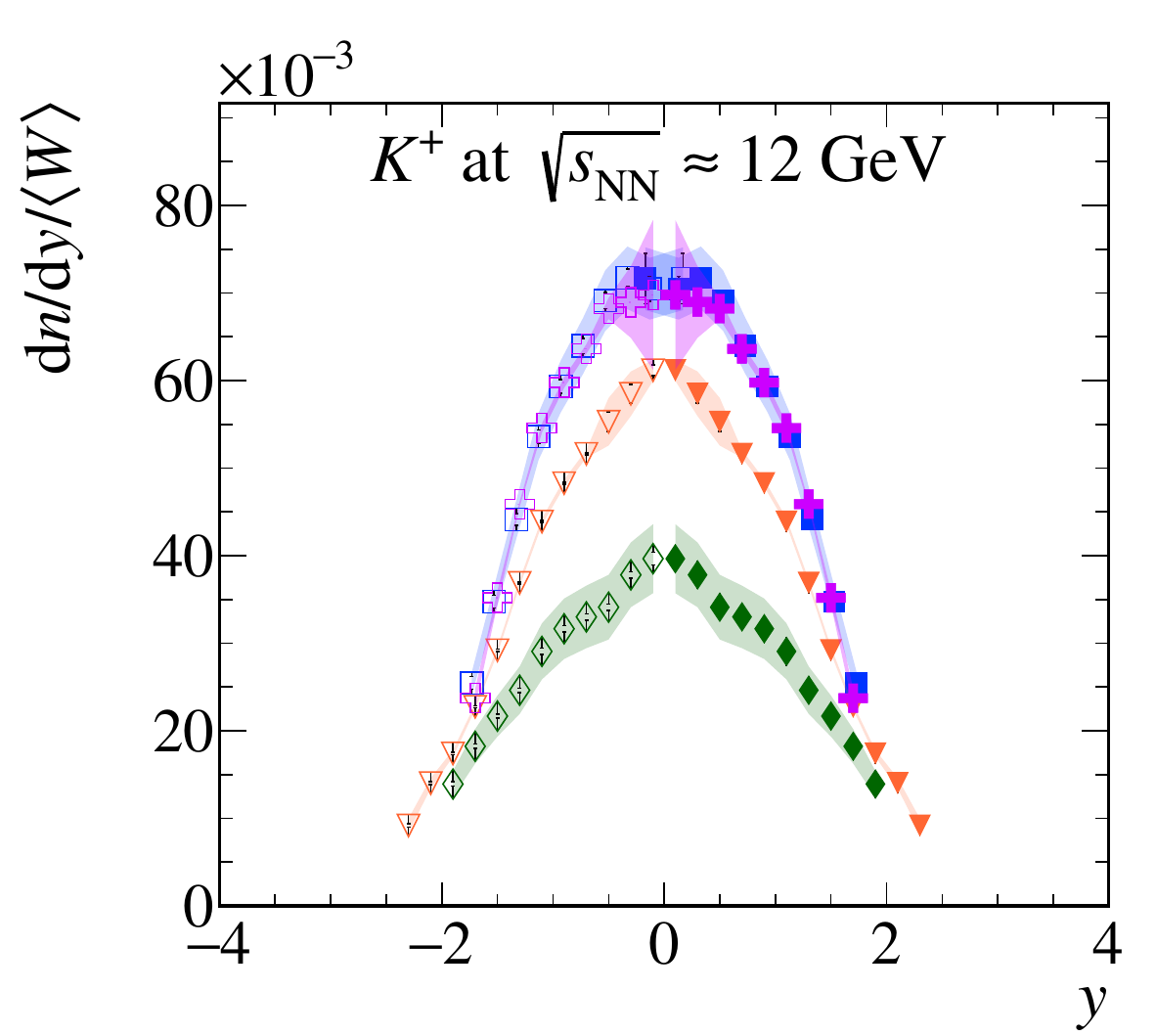}
\begin{picture}(10,10)
\put(-62,135){\tiny\bf NA61/SHINE}
\put(-229,135){\tiny\bf NA61/SHINE}
\end{picture}
  \hspace*{1cm}\includegraphics[width=0.458\textwidth]{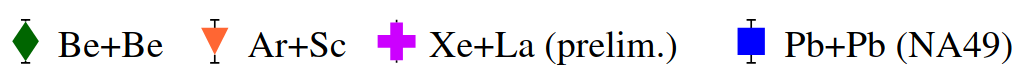}
\vspace*{-0.0cm}
  \begin{picture}(1,1)
 \put(-184, 45){\bf(a)}
\put( -17, 45){\bf(b)}
  \end{picture}
  \caption{Rapidity distributions of {\bf(a)} $\pi^-$\hspace*{-0.3mm} and {\bf(b)} $K^+$\hspace*{-0.2mm} in
  central nucleus-nucleus systems. Closed (open) symbols correspond to measured (reflected) data. Please note the non-monotonic system size dependence in panel (a).
  From Ref.~\cite{op}.}
   \label{f3}
   \vspace*{-0.2cm}
\end{figure}
\noindent
Figure~\ref{f3}~(a) shows an example set of c.m.s.~rapidity distributions of $\pi^-$\hspace*{-0.3mm} produced in central Be+Be, Ar+Sc, Xe+La, and Pb+Pb collisions at $\sqrt{s_\mathrm{NN}}\approx9$~GeV. The corresponding beam rapidity position is $y\approx2.2$, which emphasizes the very extended coverage of the measurement. Being normalized by the average number of wounded nucleons, the distributions exhibit a surprising non-monotonic system-size dependence, with the per-wounded nucleon mid-rapidity density increasing from Be+Be up to Ar+Sc, to decrease down to Xe+La and further down to Pb+Pb collisions. For comparison, charged kaon spectra exhibit a very different behavior shown in Fig.~\ref{f3} (b). A very large strangeness enhancement from Be+Be up to Ar+Sc collisions is followed by a smaller increase in Xe+La reactions, where it saturates with no further change up to central Pb+Pb collisions. 

\begin{SCfigure}[0.9][!h]
  \centering
  \includegraphics[width=0.498\textwidth]{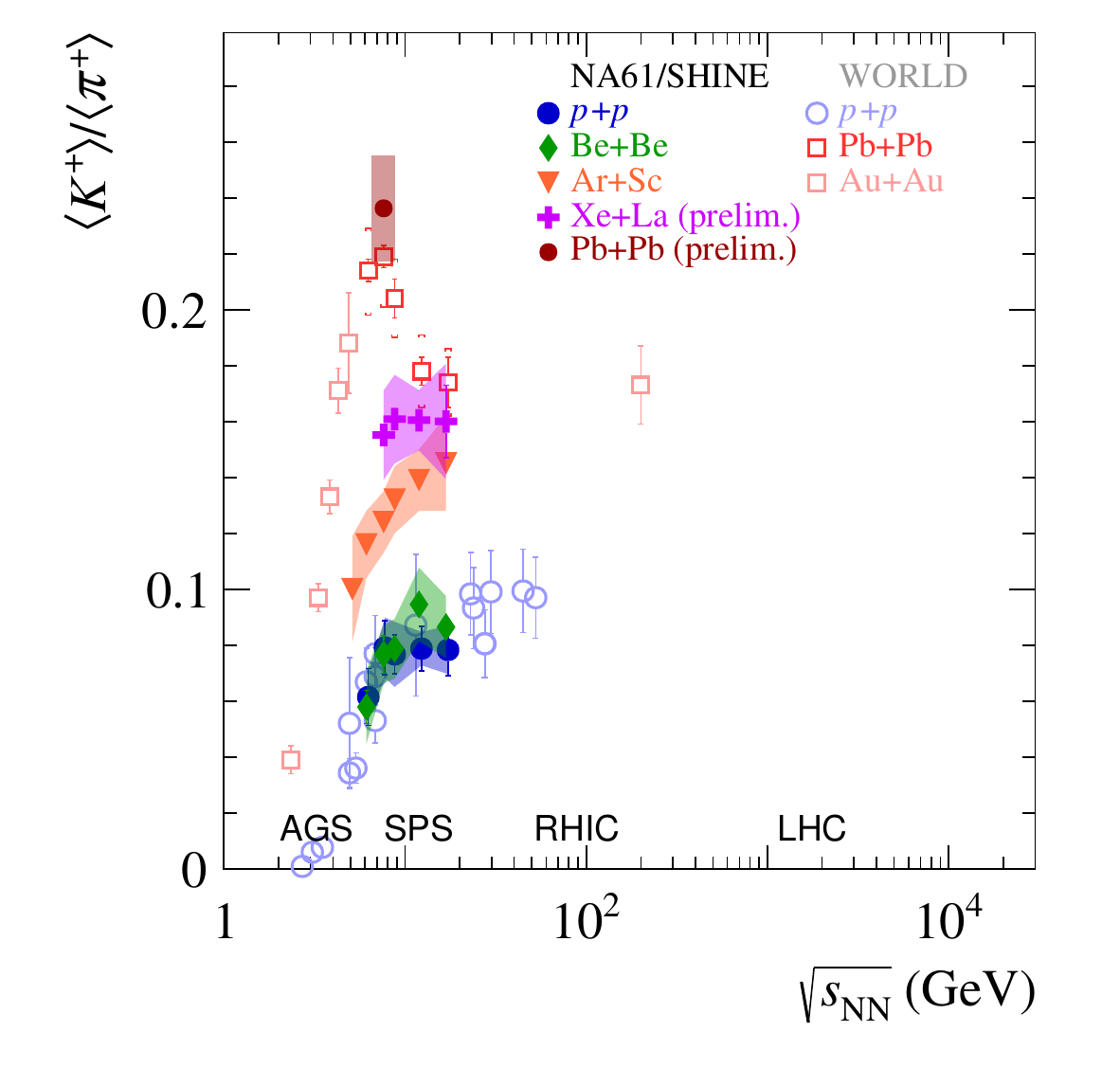}
  \caption{Energy dependence of the $K^+/\pi^+$ ratio of 4$\pi$ mul\-ti\-pli\-cities in different systems. See Ref.~\cite{arsc} for the compilation of published data. Please note that no 4$\pi$ multiplicities were found at LHC energies due to experimental limitations, but mid-rapidity $K^+/\pi^+$ ratios display a simi\-lar trend~\cite{op}. \vspace*{1cm}}
\label{f4}       
\vspace*{-0.4cm}
\end{SCfigure}
\noindent
The corresponding compilation of $\langle K^+\rangle/\langle\pi^+\rangle$ ratios of total kaon and pion multiplicities, Fig.~\ref{f4}, shows rapid changes in this ratio occurring in the specific region of $\sqrt{s_\mathrm{NN}}$ accessible to NA61/SHINE, above the SIS18/SIS100 regime but below the top RHIC/LHC region. The non-monotonic ``horn'' present in Pb+Pb/Au+Au collisions was earlier predicted~\cite{g} as a signature of the beginning of QGP creation as a function of collision energy (the onset of deconfinement). However, a characteristic, non-trivial two-dimensional dependence of this effect on system-size and energy is evident in the Figure. 
Already in Xe+La reactions (about 2/3 of the central Pb+Pb system in terms of the total number of colliding nucleons) a flattening in the $\langle K^+\rangle/\langle\pi^+\rangle$ is apparent as a function of $\sqrt{s_\mathrm{NN}}$, followed by a fully monotonic behavior for all the smaller systems.
At the present moment, the ensemble of experimental $K^+/\pi^+$ data in the SPS sector remains not understood in terms of quantitative model calculations (see Ref.~\cite{op} and references therein for more details).

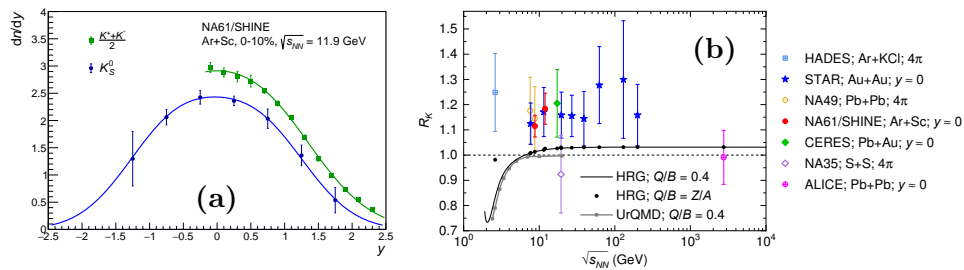
\begin{figure*}[!b]
\centering
\vspace*{-0.6cm}       
\hspace*{-0.6cm}
\includegraphics[width=0.452\textwidth]{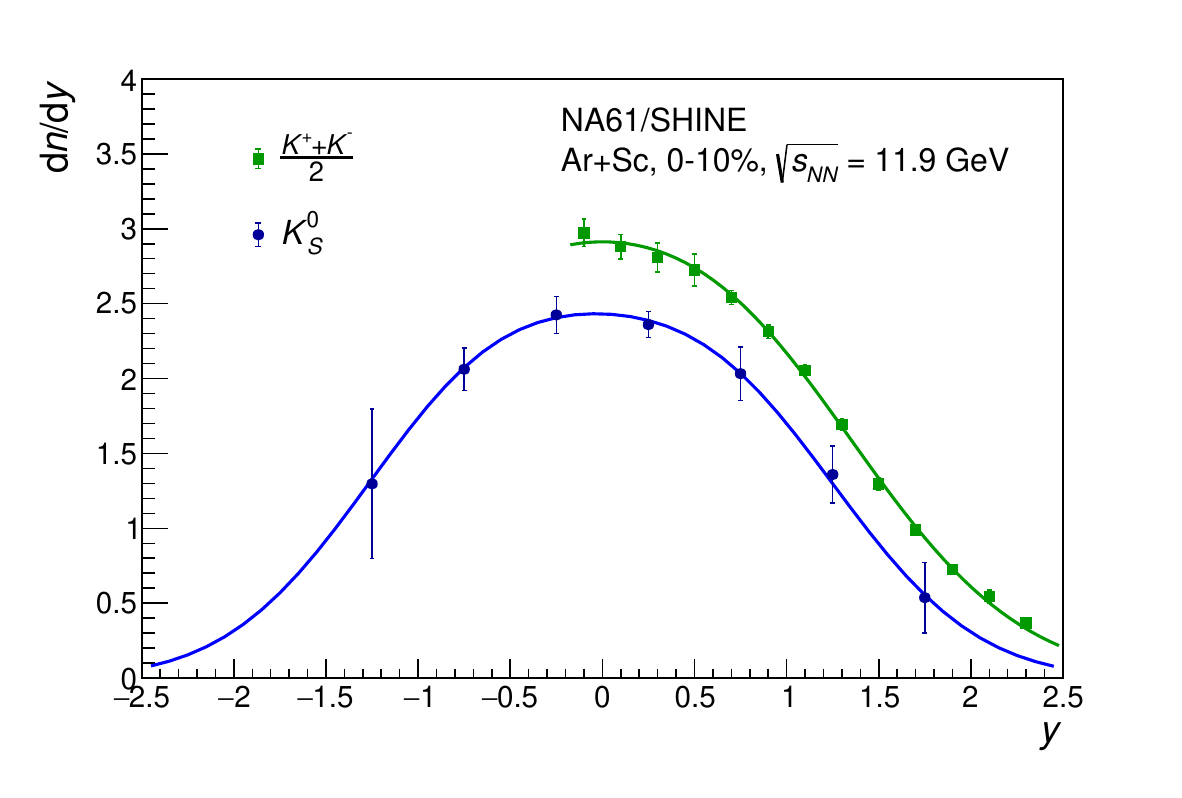}
\hspace*{-0.44cm}
\includegraphics[width=0.573\textwidth]{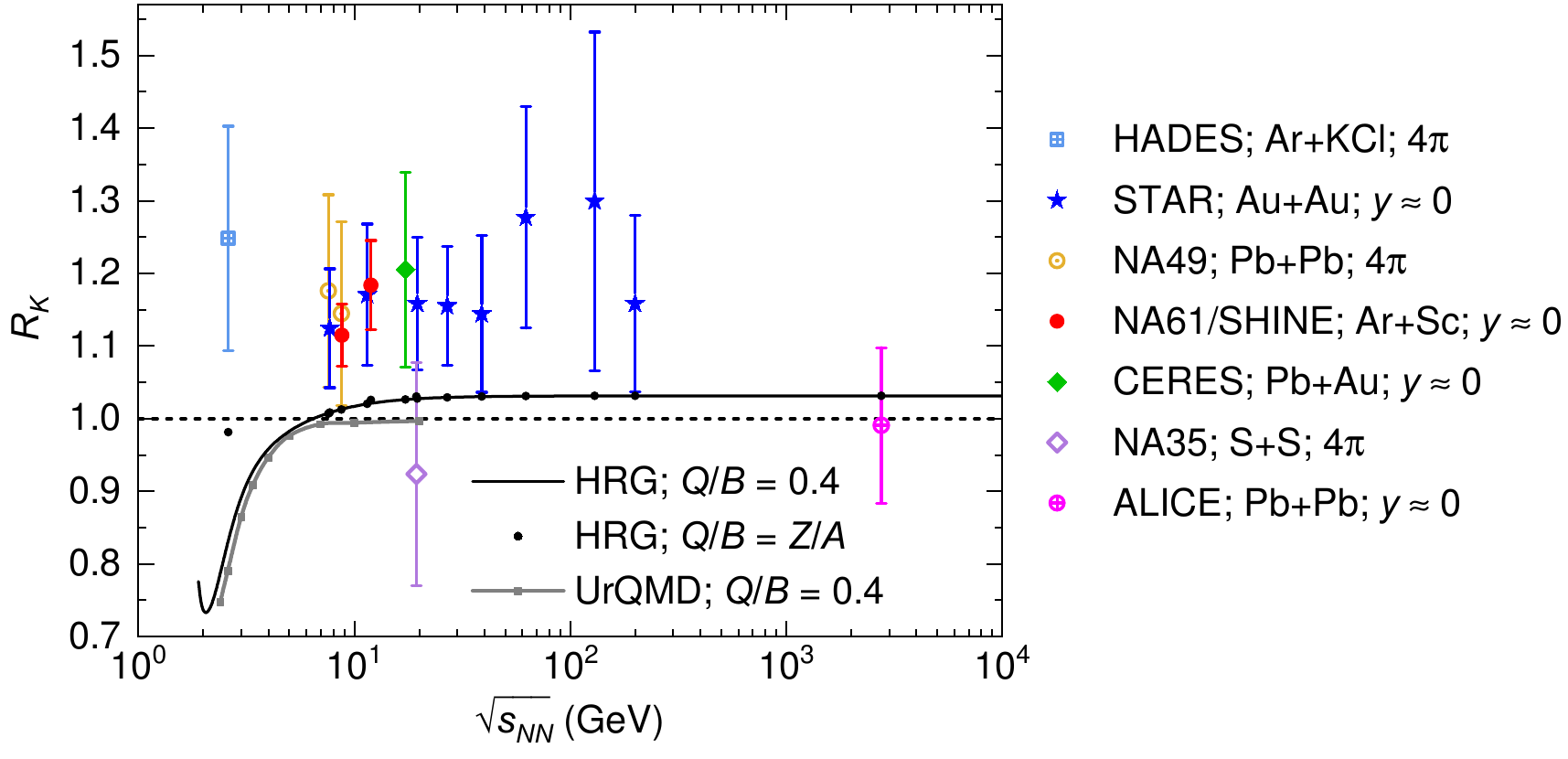}
\hspace*{-0.6cm}
  \begin{picture}(1,1)
 \put(-278,25){\bf(a)}
 \put(-92,81){\bf(b)}
\vspace*{0.3cm}       
  \end{picture}
  \caption{{\bf(a)} Comparison of rapidity spectrum of neutral kaons ($K^0_S$) with the averaged spectrum of charged kaons ($K^+$$+$$K^-$)/2 in Ar+Sc collisions. From Ref.~\cite{3}. {\bf(b)} Ratio $R_K\,$=$\,(K^+$$+$$K^-)$/(2$K^0_S$), drawn as a function of $\sqrt{s_\mathrm{NN}}$. NA61/SHINE data include $\sqrt{s_\mathrm{NN}}$=11.9~\cite{3} and 8.8~GeV~(preliminary).~The compilation of results from earlier experiments includes ratios of mid-rapidity densities as well as 4$\pi$ yields.  
    Predictions from UrQMD and HRG models are also shown; $Q/B$ is the assumed charge-to-baryon number ratio. From Ref.~\cite{3}
    (see therein for the bi\-blio\-graphy)
    with addition of the 8.8 GeV data point.
    Total uncertainties are drawn.}
\label{f5}       
\end{figure*}

\vspace*{-0.1cm}
\section{Kaon sector of multiparticle production: \\unexpectedly strong violation of isospin (flavor) symmetry}
\noindent
Another puzzling observation from NA61/SHINE is presented in Fig.~\ref{f5}~(a). Flavor symmetry being a basic feature of QCD in the limit of equal quark masses, the remnant approximate light u/d quark symmetry (isospin symmetry) implies equal production of charged and respective neutral kaons in case of colliding charge-symmetric ($Z=N\equiv A-Z$) nuclei. For the numbers of produced $K^+\approx K^0$ and $K^-\approx \overline{K}\hspace{0.1mm}^0$, this corresponds to
\begin{equation}
  ~\frac{K^{+}+K^{-}}{2}
  \approx
  \frac{K^0+\overline{K}\hspace{0.1mm}^0}{2}=K^0_S\,,
\end{equation}
where the last equality neglects the very small effect of CP violation. However, experimental data from collisions of nearly\footnote{The
Ar and Sc nuclei are not exactly charge-symmetric. The moderate excess of neutrons~(udd) over protons~(uud) would be
expected to decrease charged kaon compared to neutral kaon production, contrary to the result shown in Fig.~\ref{f5}.}charge-symmetric
Ar+Sc nuclei at $\sqrt{s_\mathrm{NN}}$=11.9 GeV do not follow this prediction, and exhibit a charged-over-neutral kaon excess reaching 18.4\,$\pm$\,6.1\% at mid-rapidity~\cite{3}. This implies an unexpectedly strong violation of isospin/flavor symmetry for kaons in multiparticle production, as demonstrated in Fig.~\ref{f5} (b). The latter puts together the NA61/SHINE mid-rapidity result at $\sqrt{s_\mathrm{NN}}$=11.9 GeV, the corresponding preliminary result at $\sqrt{s_\mathrm{NN}}$=8.8 GeV, and a compilation of charged-over-neutral kaon ratios $R_K$ obtained from published data from other experiments (see Methods part of Ref.~\cite{3} for details on the compilation). Although with considerable uncertainties, quite a consistent picture emerges of an excess of kaons containing u ($\mathrm{\bar{u}}$) valence quarks (anti-quarks) up to $\sqrt{s_\mathrm{NN}}$=200 GeV.
The ALICE data point at $\sqrt{s_\mathrm{NN}}$=2.76~TeV may indicate that the effect fades out at higher energies, but the uncertainties remain too large to allow definite conclusions. It is to be noted that a compilation of ALICE data shows that values of the charged-over-neutral kaon ratio $R_K$ at 0.9$<$$\sqrt{s_\mathrm{NN}}$$<$13~TeV remain consistent with unity independently on reaction type and collision energy, but quoted uncertainties remain sizeable, on the level of 6$-$11\%~\cite{e}. Consistently with the above, a charged-over-neutral kaon excess is also observed by NA61/SHINE in $\mathrm{\pi}^-$+C reactions at $\sqrt{s}$=17.3 and 25.7~GeV in disagreement with model predictions~\cite{m}, although any claim of violation of isospin symmetry in this case necessitates the analysis of the full charge-symmetric ensemble of the new NA61/SHINE $\mathrm{\pi}^\pm$+C data.

The comparison of experimental data with models also shown in Fig.~\ref{f5}~(b) demonstrates the failure of explaining the measured results by known effects violating isospin symmetry (see Ref.~\cite{3} for discussion of the included effects). This, of course, implies consequences for the present understanding of mechanisms of particle production in the non-perturbative sector of strong interactions. The present success in explaining these results by new theoretical ideas can be considered as moderate (author’s personal opinion). A modification of the UrQMD model allowing for a quantitative description of the charged-over-neutral kaon excess in nucleus-nucleus collisions necessitates assuming a high (3:1) u:d asymmetry in the string fragmentation process~\cite{b}. Notably, this allows for a consistent description of similar effects in $p$$+$$p$ at\hspace*{1mm}$\sqrt{s_\mathrm{NN}}$=12.3 GeV, and $e$$^+$$+$$e$$^-$ data at\hspace*{1mm}$\sqrt{s_\mathrm{NN}}$=3.050 GeV~\cite{b}, even if the two latter reactions contain also known effects that can deviate the ratio $R_K$ from unity (valence u excess, QED contribution). Also, the modified model predicts a moderate decrease of $R_K$ towards LHC energies. A discussion of theoretical ideas inspired by the NA61/SHINE observation will soon appear in Ref.~\cite{wp}.

\section{Hidden strangeness}
\noindent
With its almost pure s$\mathrm{\bar{s}}$ valence content, the $\mathrm{\phi}$(1020) meson can be considered a good tool for development of phenomenological models of the soft sector of the strong interaction. It is also believed to be a sensitive probe of the hadronic or partonic nature of the system created in the collision. The meson is supposed to behave as a strangeness-neutral particle in a purely hadronic scenario, and as a doubly-strange particle if partonic degrees of freedom are significant~\cite{pp}. The NA61/SHINE measurements in $p$$+$$p$ and central Ar+Sc collisions bring insight in both aspects, confirming the SPS energy regime ($\sqrt{s_\mathrm{NN}}\,$=\,5$-$17~GeV) as challenging for a number of models of particle production.
\begin{figure}[!h]
  \centering
  \hspace*{0cm}\includegraphics[width=0.458\textwidth,page=3]{ArSc75_043_RapiditySpectrumWithFit_MODELS.pdf}
  \hspace*{0cm}\includegraphics[width=0.458\textwidth]{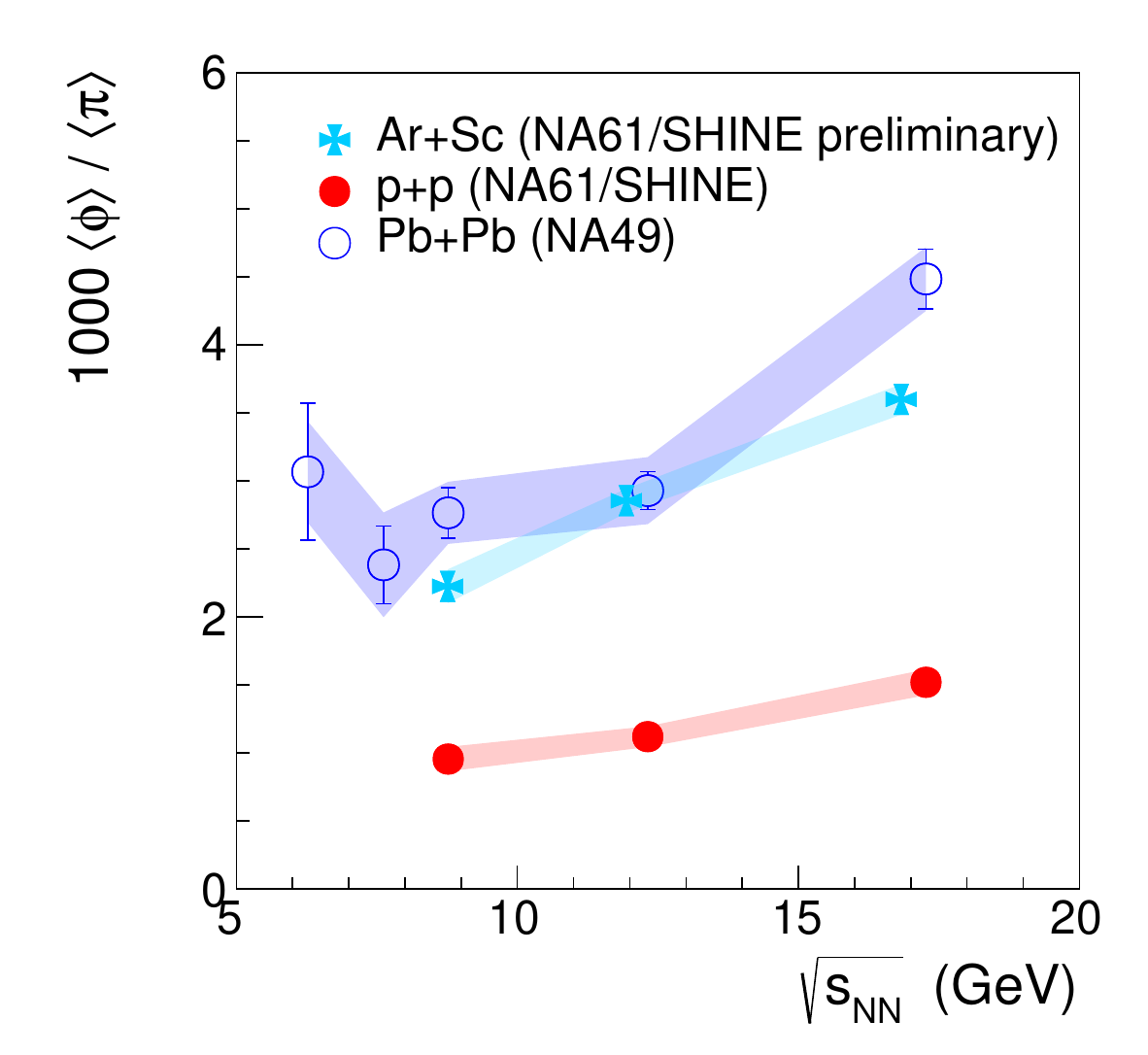}
\begin{picture}(10,10)
\put(-292,143){\tiny\bf NA61/SHINE preliminary}
\end{picture}
\vspace*{-0.0cm}
  \begin{picture}(1,1)
 \put(-221, 101){\bf(a)}
\put( -49, 33){\bf(b)}
  \end{picture}
  \caption{\hspace*{-0.2cm}{\bf(a)} Rapidity distribution of $\mathrm{\phi}$(1020) mesons produced in 0$-$10\% central Ar+Sc collisions at $\sqrt{s_\mathrm{NN}}$=11.9 GeV, in comparison to model predictions. Model calculations by S.~Veli (Technical University of Munich) and T.~Janiec (The University of Manchester). From Ref.~\cite{amar}, see therein for full bibliography. {\bf(b)} Collision energy dependence of the ratio of total $\phi$ to total pion yields ($\langle$$\pi$$\rangle$$\equiv$$\frac{3}{2}$$($$\langle$$\pi^+$$\rangle$$+$$\langle$$\pi^-$$\rangle$$)$$)$ in
    inelastic $p$$+$$p$, central Ar+Sc, and central Pb+Pb collisions. From Ref.~\cite{lr}.}
   \label{f6}
\end{figure}

\noindent
Figure~\ref{f6}~(a) shows the rapidity distribution of $\mathrm{\phi}$ mesons produced in the projectile hemisphere ($y$$>$$0$) of central Ar+Sc collisions at $\sqrt{s_\mathrm{NN}}$=11.9 GeV, put in comparison with the predictions of Pythia/Angantyr and two versions of the UrQMD model. The failure of the three model calculations to describe the experimental data is evident. Pythia/Angantyr strongly underpredicts the Ar+Sc data, which goes well in line with the underprediction of NA61/SHINE $p$$+$$p$ data by Pythia~6 in Ref.~\cite{pp}. The UrQMD and UrQMD+hydro, respectively, strongly underpredict and overpredict the data. The energy and system size dependence of hidden strangeness-over-pion yields in the SPS regime is shown in Fig.~\ref{f6}~(b). These appear strongly sensitive to nuclear effects, with a large enhancement of hidden strangeness production from $p$$+$$p$ to Ar+Sc collisions, and a much smaller further enhancement (if any) in central Pb+Pb reactions. As shown in Ref.~\cite{lr}, the enhancement effect is comparable, or slightly higher, than for kaons suggesting a role of partonic degrees of freedom in the production process.

\vspace*{-0.2cm}
\section{First direct measurement of open charm production in nucleus-nucleus collisions at SPS energies}

\begin{SCfigure}[0.65][!h]
  \centering
  \vspace*{-1.0cm}
  \hspace*{-0.3cm}
  \includegraphics[width=0.598\textwidth]{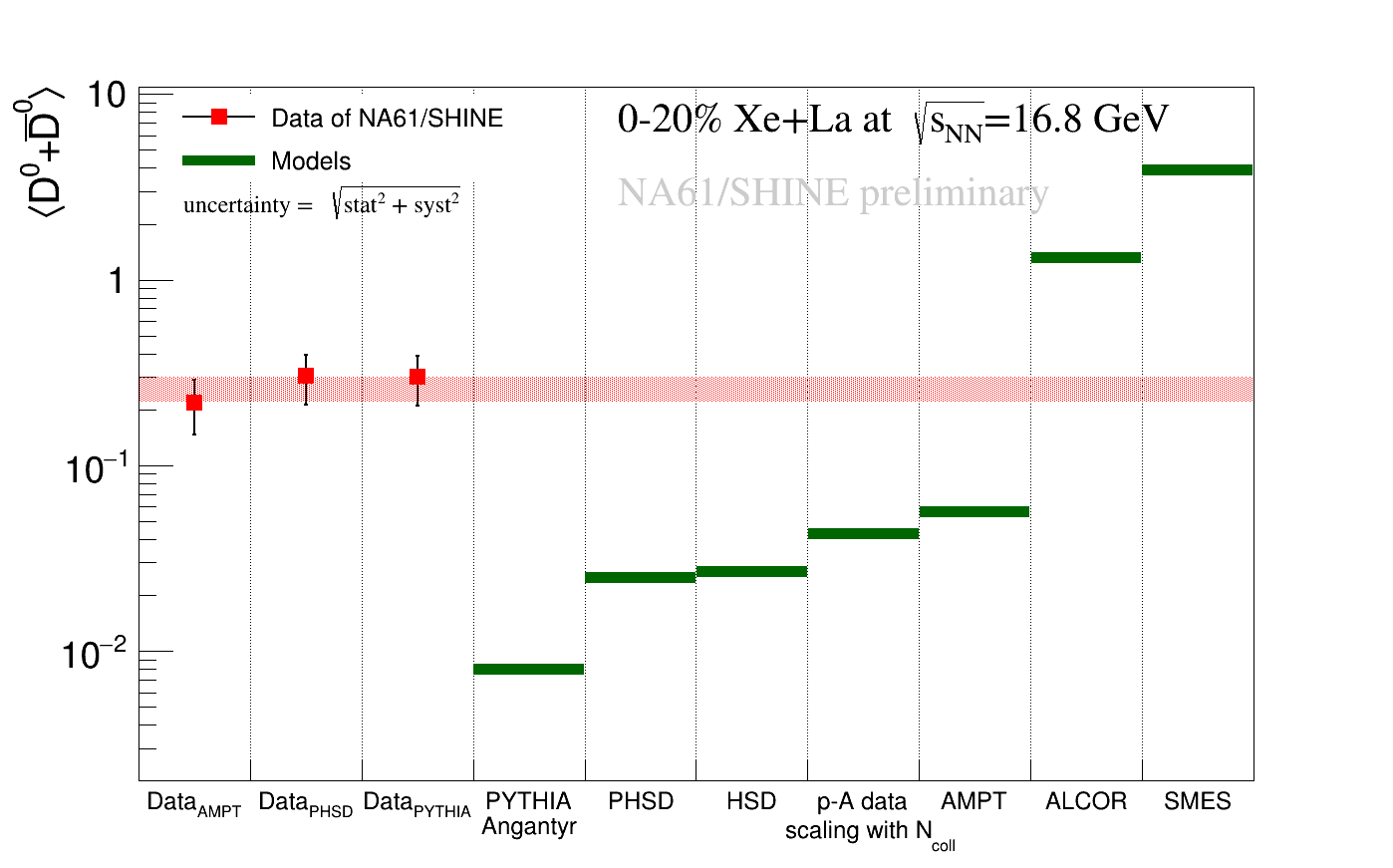}
  \caption{The open charm 4$\pi$ yield $\langle D^0$$+$$\overline{D}{\hspace{0.1mm}}^0\rangle$ in central Xe+La collisions at $\sqrt{s_\mathrm{NN}}$=16.8 GeV, put in comparison to predictions of several theoretical models. The red band gives the (model-related) extrapolation uncertainty of the 4$\pi$ yield. For numerical values and a full bibliography of the theoretical models, see Ref.~\cite{am}.\vspace*{-1.2cm}}
  \label{f7}       
\end{SCfigure}

\noindent
The mechanism of c, $\mathrm{\bar{c}}$ quark and anti-quark production in nucleus-nucleus collisions remains poorly known in the SPS regime of relatively low collision energies. To illustrate this, Fig.~\ref{f7} gives a compilation of model predictions obtained for\hspace*{1mm}$\sqrt{s_\mathrm{NN}}$=16.8 GeV, ranging from 0.008 up to 4 charmed $D^0$ and $\overline{D}\hspace*{0.1mm}^0$ mesons produced per central Xe+La event. NA61/SHINE pursues a research program aimed at direct measurements of $D^0$ and $\overline{D}\hspace*{0.1mm}^0$ yields
and
kinematic spectra in Pb+Pb collisions.
As the first result of this program obtained with a prototype silicon vertex detector, a low-statistics measurement of 170\,$\pm$\,28 $D^0$ and $\overline{D}\hspace*{0.1mm}^0$ mesons has been obtained.
The $D^0$$+$$\overline{D}\hspace*{0.1mm}^0$ yield
in acceptance was extrapolated to the total 4$\mathrm{\pi}$ yield using three different models to estimate the corresponding model-related uncertainty. The result is presented in Fig.~\ref{f7}. As evident from the figure, and in spite of its sizable statistical, systematic and extrapolation-related uncertainties, this first direct measurement of $D^0$, $\overline{D}\hspace*{0.1mm}^0$ production in $A$$+$$A$ collisions at the SPS provides a very strong constraint on the largely differing models.
%
As such, it gives the first chance of identifying the scenario underlying open charm production at this energy.
%
  \vspace*{0.1cm}
\vspace*{0.6cm}

\section{Summary and conclusions}

The NA61/SHINE SPS energy regime
($5.1\hspace*{-2mm}<\hspace*{-2mm}\sqrt{s_\mathrm{NN}}\hspace*{-2mm}<\hspace*{-2mm}16.8/27.4$~GeV)
occupies an intermediate position between that of the announced FAIR SIS100 heavy ion program
($2.7\hspace*{-2mm}<\hspace*{-2mm}\sqrt{s_\mathrm{NN}}\hspace*{-2mm}<\hspace*{-2mm}4.9$~GeV)
and the LHC regime
($0.9\hspace*{-1mm}<\hspace*{-1mm}\sqrt{s_\mathrm{NN}}\hspace*{-1mm}<\hspace*{-1mm}14$~TeV).
In this context, the fact that several experimental observations related to strange\-ness production, violation of u/d flavor (isospin) symmetry, and charm production remain poorly understood by theory and meet largely differing theoretical predictions is to be seen with concern and calls for further studies.

The present short summary of NA61/SHINE results is supplemented with a more detailed account on baryon and meson spectra including strange\-ness production~\cite{op} as well as studies of correlations and fluctuations in view of the up to now unsuccessful search for the QCD critical point~\cite{nd}.

The Author warmly thanks the Organizers of an excellent Epiphany Conference in 2026. This work was
supported by the Polish Ministry of Science and Higher Education (2025/WK/05).



\end{document}